\documentclass[sigconf, nonacm]{acmart}
\usepackage[]{hyperref}
\usepackage{mathtools}
\usepackage{physics}
\usepackage[normalem]{ulem}

\newcommand{\ham}{\mathcal{H}}
\newcommand{\revision}{\textcolor{black}}

\AtBeginDocument{%
  }

\begin{document}

\title{Variational Quantum Algorithms in the era of \\
Early Fault Tolerance}


\author{Siddharth Dangwal}
\affiliation{%
  \authornote{
Correspondence: siddharthdangwal@uchicago.edu}
  \institution{University of Chicago}
  \country{USA}
}
\author{Suhas Vittal}
\affiliation{%
  \institution{Georgia Institute of Technology}
  \country{USA}
}

\author{Lennart Maximillian Seifert}
\affiliation{%
  \institution{University of Chicago}
  \country{USA}
}

\author{Frederic T. Chong}
\affiliation{%
  \institution{University of Chicago}
  \country{USA}
}

\author{Gokul Subramanian Ravi}
\affiliation{%
  \institution{University of Michigan}
  \country{USA}
}

\begin{abstract}
Quantum computing roadmaps predict the availability of 10,000-qubit devices within the next 3–5 years. With projected two-qubit error rates of 0.1\%, these systems will enable certain operations under quantum error correction (QEC) using lightweight codes, offering significantly improved fidelities compared to the NISQ era.

However, the high qubit cost of QEC codes like the surface code (especially at near-threshold physical error rates) limits the error correction capabilities of these devices. In this emerging era of Early Fault Tolerance (EFT), it will be essential to use QEC resources efficiently and focus on applications that derive the greatest benefit.

In this work, we investigate the implementation of Variational Quantum Algorithms in the EFT regime (EFT-VQA). We introduce partial error correction (pQEC), a strategy that error-corrects Clifford operations while performing \(R_z\) rotations via magic state injection instead of the more expensive T-state distillation. Our results show that pQEC can improve VQA fidelities by \revision{9.27x} over standard approaches. Furthermore, we propose architectural optimizations that reduce circuit latency by $\sim$ 2x, and achieve qubit packing efficiency of $66\%$ in the EFT regime.

\end{abstract}

\maketitle

\section{Introduction}\label{sec:introduction}

Quantum computers have the potential to revolutionize fields like chemistry~\cite{kandala2017hardware}, optimization~\cite{moll2018quantum}, and machine learning~\cite{biamonte2017quantum} by leveraging superposition, interference and entanglement to tackle problems exponentially hard for classical systems. However, current quantum systems face major challenges in scaling and fidelity that prevent them from solving classically intractable problems.

Quantum computing spans two paradigms: near-term Noisy Intermediate-Scale Quantum Computing (NISQ) and long-term Fault-Tolerant Quantum Computing (FTQC). Currently, quantum computing centers around NISQ, with devices that are error-prone and limited in qubit numbers~\cite{preskill2018quantum}, facing challenges like limited coherence, SPAM errors, gate errors and crosstalk~\cite{ravi2022quantum}. While NISQ devices cannot run large-scale programs, they hold potential for quantum advantage in tasks like Variational Quantum Algorithms (VQAs). 
Without quantum error correction (QEC), NISQ systems rely on error mitigation~\cite{czarnik2020error,Rosenberg2021,barron2020measurement,botelho2021error,wang2021error,takagi2021fundamental,temme2017error,li2017efficient,giurgica2020digital,ding2020systematic,smith2021error}, which has moved the field closer to practical utility~\cite{kim2023evidence}, though significant progress is still needed for real-world applications. In the long term, the goal is to achieve FTQC, which requires millions of qubits with low error rates and advanced QEC codes~\cite{Fowler_2012} to reach logical error rates $\sim 10^{-15}$~\cite{Gidney_2021}, enabling large-scale algorithms like quantum phase estimation (QPE), Shor's Factoring~\cite{Shor_1997}, Grover's Search~\cite{Grover96afast} and complex simulations. However, large progress across the quantum computing stack is needed for FTQC to reach its full potential. 

\begin{figure}[htbp]
    \centering
    \includegraphics[scale=1]{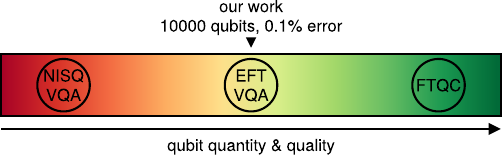}
    \caption{Variational quantum algorithms are particularly well suited to benefit from the capabilities of intermediate-scale quantum devices in the era of Early Fault Tolerance (EFT).}
    \label{fig:eftvqa}
\end{figure}

The divide between NISQ and FTQC paradigms is well known, but efforts to bridge this gap remain limited. Experimental research has focused on NISQ systems, while theoretical work targets FTQC, with small overlap between the two. This separation is evident in the architecture and systems community, where studies on NISQ~\cite{das2021adapt, tannu2019mitigating, ravi2022quancorde, murali2019noise} and FTQC~\cite{vittal2024flag, lin2024codesign, viszlai2023architecture} have progressed independently.
As skepticism about NISQ's potential grows and cautious optimism for FTQC remains, a critical question arises:
\textit{Can we effectively transition quantum computing concepts from NISQ devices to FTQC systems and identify applications that can leverage these techniques during the intermediate term of ``Early Fault Tolerance'' (EFT)?}  
We define EFT as the era with quantum systems featuring  $\sim \! 10000$ qubits and physical error rates $\sim \! 10^{-3}$, a realistic expectation for the next decade given industry roadmaps ~\cite{google_roadmap, iroadmap_2}. 
Addressing this question will require diverse ideas, and this paper focuses on one promising domain: Variational Quantum Algorithms (VQAs)~\cite{peruzzo2014variational,farhi14}.

While VQAs are known to show some noise resilience, typically the noise on NISQ machines remains too high for VQAs to achieve the accuracy needed to compete with classical techniques in many large-scale, real-world applications, even with advanced error mitigation strategies~\cite{zne1,zne3,pokharel2018demonstration,QISMET_Ravi2022}. For example, estimating the ground state energy of molecules requires an infidelity of less than $1.6 \times 10^{-3}$ Hartree, known as ``chemical accuracy''~\cite{peterson2012chemical}. However, for molecules of practical interest, NISQ VQA fidelity is expected to be orders of magnitude worse than classical methods. Given this large gap between NISQ VQA performance and real-world requirements, it is essential to start looking beyond the NISQ paradigm for VQAs.
In this work, we propose early fault-tolerant VQA (EFT-VQA) as a promising application for the EFT era. EFT-VQA can scale to larger problem sizes and achieve greater accuracy than typical NISQ VQA while requiring less QEC than full FTQC applications.

Several challenges must be overcome to make EFT-VQA practical, and this work addresses many fundamental questions. An important challenge stems from the fact that VQAs contain many rotation gates with tunable angles in \([0, 2\pi]\). The logical gate set often used for FTQC is Clifford+T \cite{ross2016optimal}, where the non-Clifford T gate is more resource-intensive to execute than Cliffords (see Section \ref{subsec:distillation}). If EFT-VQA uses this gate set, the VQA rotation gates will need to be decomposed into Clifford+T gates, requiring  hundreds of T gates per rotation for reasonable accuracy~\cite{dawson2005solovaykitaev, ross2016optimal}.

Directly porting NISQ VQA ansatzes and FTQC synthesis procedures to the EFT regime would thus result in EFT-VQA circuits requiring hundreds of thousands of T gates per iteration execution, even for medium-sized tasks. Such T-heavy circuits pose significant challenges in the EFT era, where the execution of T gates is constrained by fidelity and resource demands (we refer to  Sections \ref{subsec:drawbacks} and \ref{sec:pqec} for details). Specifically, the resulting high circuit depth can adversely affect fidelity, especially when error correction is limited. Moreover, the volume of T gates necessitates substantial qubit resource allocation for the costly process of magic state distillation \cite{knill2004fault, bravyi2005universal}, which generates high-quality T states. This resource demand becomes a critical bottleneck when qubits remain relatively constrained in the EFT era .  

Recognizing the challenges of implementing EFT-VQA with the Clifford+T gate set, we explore an alternative approach called \textit{partial quantum error correction (pQEC)} \cite{akahoshi2024partially} and adapt it to VQAs. This method replaces the traditional Clifford+T gate set with a Clifford+$R_z(\theta)$ gateset and substitutes magic state distillation with an efficient $R_z(\theta)$ gate injection process~\cite{lao2022magic, akahoshi2024partially}. This approach achieves $R_z(\theta)$ error rates comparable to those in the NISQ regime, while all other gates benefit from the significantly lower error rates enabled by conventional QEC. 
By avoiding the long gate sequences generated by Clifford+T decomposition, the overall number of $R_z(\theta)$ gates is low, which limits the impact of the higher error rates associated with the $R_z(\theta)$ gates. This results in a balanced and practical solution for EFT-VQA, making pQEC an ideal compromise for the challenges of the EFT era. 

\textbf{The key contributions and insights of our work are:}
\begin{enumerate}
\item This work is among the first to bridge the gap between NISQ and long-term FTQC by thoroughly investigating the regime of early fault tolerance (EFT).  
\item To our knowledge, this is the first study to propose modifications to  standard NISQ VQA, enabling significant benefits in the EFT era and paving the way for practical quantum advantages in VQA applications.  
\item To facilitate efficient VQA execution in the EFT era, we explore and adapt ideas of \textit{partial quantum error correction (pQEC)}. It leverages lightweight surface codes to correct errors in Clifford gates and uses \textit{magic state injection} for non-Clifford \(R_{z}(\theta)\) gates. Our results show that this strategy for VQA outperforms NISQ execution by \revision{$9.27$x} on average and up to 257.54x, even for VQA ansatz explicitly tailored for the NISQ regime. 
\item We also demonstrate that achieving the same VQA fidelity as the pQEC method with a Clifford+T gateset would require T gates of much higher fidelity than those currently produced by magic state factories within EFT constraints. In numerous cases, achieving parity with pQEC would demand T gates with a 0\% error rate --- an unachievable standard.  
\item To further optimize EFT-VQA execution, we propose several architectural enhancements, including:  \textit{efficient qubit layouts}, a \textit{patch shuffling mechanism} to streamline non-Clifford injection, a \textit{layout-aware ansatz} designed for higher efficiency in VQA execution,  and in-depth circuit-level cost analysis to identify and mitigate key overheads. These optimizations reduce circuit execution overheads like space, time, and spacetime volume.
\end{enumerate}

\section{Background and motivation}\label{section:background}

\begin{figure*}[h!]
    \centering
    \includegraphics[width=0.8\linewidth]{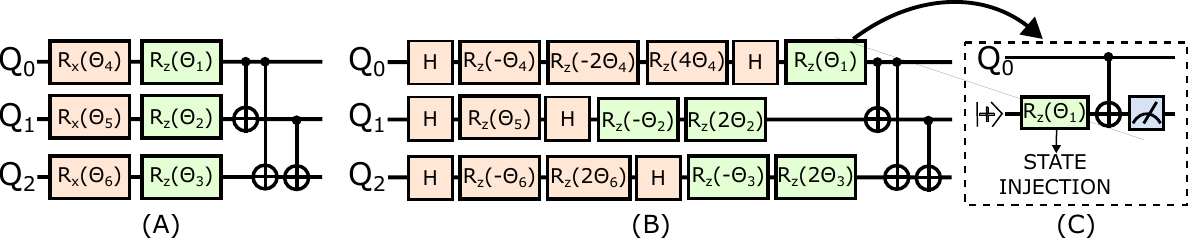}
    \caption{(A) A typical VQA circuit. (B) The effective circuit at runtime. Ancilla qubits are omitted for clarity. (C) The circuit used to implement an $R_{z}(\theta_1)$ rotation. An ancilla qubit, initialized in the $\ket{+}$ state is injected with a magic state which is later consumed by the data qubit to implement a logical rotation.}
    \label{fig:rotation}
\end{figure*}

\subsection{Achieving utility with VQAs}\label{sec:vqa}

Despite errors on near-term quantum devices, some algorithms, like variational quantum algorithms (VQAs), may allow noisy quantum computers to outperform classical ones. A VQA consists of a parameterized circuit (``ansatz'') and a Hermitian operator (``Hamiltonian'') aimed at minimizing a scalar loss (``energy'') measured on quantum hardware. Classical optimizers, such as SPSA and Nelder-Mead~\cite{TILLY20221}, adjust the ansatz parameters to achieve this. VQAs exhibit Optimal Parameter Resilience (OPR)~\cite{wang2021can}, where parameters that minimize loss on noisy hardware often do so on noiseless devices, making them theoretically noise-robust. However, NISQ noise is so high that error reduction is essential for OPR on practical problems.  

The Variational Quantum Eigensolver (VQE)~\cite{peruzzo2014variational}, a popular VQA, estimates the minimum eigenvalue of a Hamiltonian and is widely applied in physics and chemistry, such as finding molecular dissociation energies and Ising Model energies. This work focuses on VQEs but extends to other VQAs like QAOA~\cite{farhi2014quantum} and QML~\cite{biamonte2017quantum}. 

Near-term error reduction primarily relies on error mitigation. Despite numerous techniques proposed for NISQ hardware~\cite{dangwal2023varsaw, ravi2021vaqem, ravi2022cafqa, ravi2022qismet, li2022paulihedral, jin2024tetris, seifert2024clapton, giurgica2020digital}, NISQ noise remains too high for VQAs to be practically usable. Many VQA applications, while tolerating some error, require high fidelity (e.g., >99\%) to outperform classical methods. While novel error mitigation strategies are essential, it is increasingly clear that incorporating (limited or partial) quantum error correction is the only viable path to significantly reduce noise. 




\subsection{Error correction and surface codes}\label{sec:qec}
The core concept of \textit{quantum error correction (QEC)} is to encode a logical qubit into many \textit{data qubits}, which are continuously monitored for errors via \textit{ancilla qubits}. Together, these physical qubits form the logical qubit.
A \textit{quantum error correction code (QECC)} describes the specific encoding of the logical qubits and is characterized by its \textit{code distance $d$}, indicating the number of detectable ($d-1$) and correctable (\(\lfloor\frac{d-1}{2}\rfloor\)) errors. Some codes handle different error types, such as bit flips ($X$ errors) and phase flips ($Z$ errors), with separate distances ($d_x$ and $d_z$). Codes may also have a ``temporal distance'' ($d_m$) for resilience to measurement errors~\cite{litinski2019magic}.

\begin{figure}[h!]
    \centering
    \includegraphics[width=\linewidth]{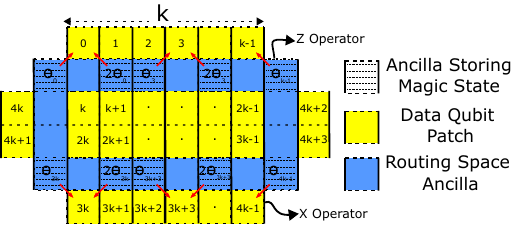}
    \caption{The layout of logical qubit patches. The parameter $k$ controls the number of allowed logical qubits. The data qubits are numbered from 0 to $4k+3$. The blue space corresponds to ancilla qubits which can be used for routing or storing $R_{z}(\theta)$ magic states. Different magic states can be stored which enables us to implement multiple $R_{z}$ rotations in parallel.}
    \label{fig:tile_layout}
\end{figure}

A prominent QECC is the \textit{surface code}, which requires $d^2$ data qubits and $d^2-1$ ancilla qubits for a code of distance $d$ \cite{Fowler_2012}. 
A qubit encoded by the surface code is often referred to as a ``patch''. Multi-qubit gates like CNOTs are executed via \textit{lattice surgery} \cite{fowler2018low}, where logical data qubits are placed among logical ancilla qubits to enable interactions. For operations between distant logical qubits, contiguous ancilla channels are reserved, with exclusive use ensuring no overlap. Allocating more ancilla qubits reduces execution time. Figure~\ref{fig:tile_layout} illustrates such a layout, where yellow patches represent logical data qubits and green and blue patches denote logical ancilla qubits. Section \ref{subsec:good_layout} will further discuss this layout.

\subsection{Universal quantum gate sets}
\label{subsec:universal_gate_set}
A universal quantum gate set allows for the approximation of any unitary operation to arbitrary precision. One of the most well-known universal gate sets for fault-tolerant quantum computing (FTQC) is the Clifford+T set. While Clifford gates (such as Hadamard, CNOT and the Pauli gates) play a crucial role in quantum circuits, they are not universal on their own. The addition of a non-Clifford gate, in this case the T gate, addresses this limitation, enabling the construction of arbitrary unitary operations. Although Clifford+T is a discrete, finite gate set, it can still approximate any unitary to arbitrary precision \cite{kliuchnikov2012fast, giles2013exact}. However, achieving high precision requires decomposing non-Clifford quantum gates into long sequences of Clifford+T, using techniques like Gridsynth~\cite{ross2016optimal}, where greater precision leads to longer gate sequences. This results in significant overhead, especially in the Early Fault Tolerant (EFT) era, as discussed in Section \ref{subsec:drawbacks}.

To mitigate these overheads, another gate set worth considering is Clifford + \( R_z(\theta) \), where \( R_z(\theta) \) represents arbitrary continuous rotations around the z-axis. This gate set is especially well suited for VQAs, since all non-Clifford gates in VQAs are $R_{z}(\theta)$ gates typically. The advantage of this gate set is that operations requiring high-precision rotations can be implemented without dramatically increasing the number of gates. However, generating and executing high-precision \( R_z(\theta) \) gates presents its own challenges, which must also be taken into account. These challenges will become evident in the subsequent section.

\subsection{Magic state distillation}\label{subsec:distillation}

The challenges associated with executing T or \( R_z(\theta) \) gates in the FTQC paradigm stem from the Eastin-Knill theorem \cite{eastin2009restrictions}, which imposes a fundamental limitation on quantum error correction (QEC) codes. It states that no code can implement a fault-tolerant universal gate set natively, meaning no code can correct errors in all gates required for universality. For instance, the surface code can correct errors in Clifford operations but not in non-Clifford gates like T or \( R_z(\theta) \). 

To address non-Clifford gates, especially the T gate, a procedure known as \textit{magic state distillation} \cite{knill2004fault, bravyi2005universal, litinski2019magic} is typically employed, which is explored here in the context of T gates (hence also called \textit{T state distillation}). This process produces a high-fidelity T state $\ket{T} = \ket{0} + \exp(i\frac{\pi}{4})\ket{1}$ with additional qubit-cycle overhead. The resulting low-error T states are then consumed by a qubit in any arbitrary state to implement a T gate. 


The implementation of T state distillation involves starting with a few surface code patches in the state \( \ket{+} = (\ket{0} + \ket{1})/\sqrt{2} \) and applying a sequence of logical rotations to them. The number of surface code patches required depends on the specific distillation protocol (``magic state factories'') being used \cite{litinski2019magic}. A subset of qubits is measured, while the unmeasured qubits have a T state encoded in them. The measurement output serves as an error signature, indicating whether any errors occurred during the logical rotations. If the error signature is positive, meaning an error was detected, the T states are discarded and the process is repeated. If no errors are detected, the output consists of high-fidelity \textit{distilled T states}. However, it is important to note that the error signature cannot detect certain types of errors, which is why the resulting T states, while having a very low error rate, are not entirely error-free.

T state distillation protocols are characterized by the number of surface code patches required as input and the number of distilled T states produced as output, along with the parameters \(d_X\), \(d_Z\), and \(d_m\) of the surface code patches used in the magic state factory~\cite{litinski2019magic}. For instance, the commonly used (15-to-1)\(_{7, 3, 3}\) factory requires 15 input patches to produce one distilled T state, with \(d_X = 7\), \(d_Z = 3\), and \(d_m = 3\). 
The error rate of the output T state improves as the factory’s space and time overheads increase. These overheads can be adjusted by increasing \(d_X\), \(d_Z\) (space), or \(d_m\) (time). Additionally, factories with more input patches or higher output production inherently require greater resources. For example, under the same \(d_X\), \(d_Z\), and \(d_m\) values, a (20-to-4) factory incurs higher space and time costs compared to a (15-to-1) factory.

\subsection{Drawbacks of distillation-based Clifford+T execution for EFT}\label{subsec:drawbacks}

As mentioned earlier, the generation and execution of high-precision T gates is considerably more resource-intensive than Clifford gates in the FTQC paradigm, making the Clifford+T gate set challenging to implement. Below, we outline some of these challenges:

\begin{enumerate}
    \item \textbf{Long Clifford+T decomposition}: The Clifford+T decomposition of most non-Clifford operations involves hundreds of gates, which significantly increases the depth and gate count of quantum circuits. For instance, in a 20-qubit Variational Quantum Eigensolver (VQE) circuit, a \( 10^{-6} \) precision Gridsynth~\cite{ross2016optimal} decomposition can increase the circuit depth by a factor of 7 and the gate count by 20 times. This severe growth in complexity poses a significant challenge for fault-tolerant quantum computing.
    \item \textbf{High distillation overhead}: Magic state distillation introduces substantial overhead. The simplest distillation protocol (15-to-1)$_{7, 3, 3}$ requires 810 physical qubits and takes 22 clock cycles to produce one T state for a physical error rate of \( 10^{-3} \), resulting in T states with an error rate of \( 5.4 \times 10^{-4} \) \cite{litinski2019magic}. This relatively small reduction in error comes at a high cost, especially when paired with the increased gate count from Gridsynth decompositions. For a quantum computer with a physical qubit budget of 10000 qubits, over 8\% of the qubits are dedicated to distillation, limiting the scale of the programs that can be executed. Schemes such as (15-to-1)$_{17, 7, 7}$, which produce higher fidelity T states $(4.5\times10^{-8})$, require up to 46\% of physical qubits and 42 clock cycles to generate a single T state. Additionally, the extended time needed for distillation increases vulnerability to memory errors, as qubits must wait longer for a magic state. This can be mitigated by adding more magic state factories, but doing so further reduces the qubits available for logical operations.
\end{enumerate}

These overheads are particularly relevant for EFT-VQAs, as the VQA ansatz includes numerous tunable \( R_z(\theta) \) gates, with the precision required for these gates potentially varying throughout the algorithm's execution. As a result, synthesizing the EFT-VQA ansatz using Clifford+T gates can lead to significant circuit overheads.

\subsection{Alternate strategy: magic state injection}\label{subsec:injection}
From the prior discussion, it is evident that distillation-based Clifford+T execution comes with significant overheads, making it essential to explore alternative methods. One promising approach which has gained recent attention is magic state injection \cite{lao2022magic}. In this method, we begin with a surface code patch in the \( \ket{0} \) or \( \ket{+} \) state and apply a physical T gate directly to one of the data qubits in the patch. This is followed by measuring stabilizer qubits in a specific sequence, which prepares the T state at the logical level. Unlike distillation, which involves logical rotations and measurements, magic state injection relies on physical operations, meaning that (a) this is a  cheaper process than distillation and (b) the error rate is closely tied to the physical error rate (\( p_{phys} \)). 

A particularly interesting aspect of magic state injection is that it can be extended to inject any arbitrary non-Clifford state, not just the T state. Lao et al. \cite{lao2022magic} propose a method to reduce the error rate for injecting an arbitrary \( R_z(\theta) \) magic state into a surface code patch. Their technique achieves an error rate of \( \frac{23p_{phys}}{30} \) for a CNOT error rate of \( p_{phys} \), and initialization and single-qubit error rates of \( \frac{p_{phys}}{10} \). The ability to inject arbitrary \( R_z(\theta) \) states is particularly valuable for applications like EFT-VQAs, which contain many tunable \( R_z(\theta) \) gates. While the error rates from \( R_z(\theta) \) injection are not as low as those achievable through distillation, this method is well-suited for intermediate applications such as EFT quantum algorithms, including EFT-VQAs. These applications demand higher execution fidelity than noisy intermediate-scale quantum (NISQ) systems but do not require the extreme precision needed for long-term FTQC.

Note that the fidelity of an \( R_z(\theta) \) state can be improved by post-selecting over multiple (more than two) rounds or ``pre-distillation'' \cite{campbell2018magic}, however, this comes at additional overhead. In this work we do not employ these techniques. The cost vs benefit tradeoffs for these techniques are worthy of exploration in future work.


\section{Partial quantum error correction (pQEC) for EFT-VQA}\label{sec:pqec}

\subsection{Partial correction via state injection}

\begin{figure}[h!]
    \centering
    \includegraphics[width=\linewidth]{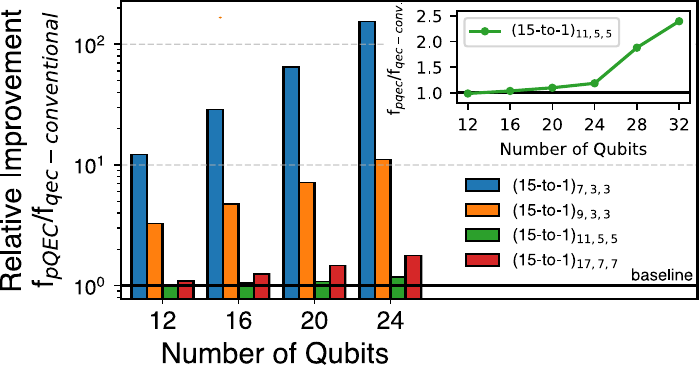}
    \caption{Relative fidelity improvement for VQAs when executed using pQEC compared to \texttt{qec-conventional}. For all cases, pQEC does as well as or outperforms \texttt{qec-conventional}.} 
    \label{fig:pQEC_vs_qec_conventional}
\end{figure}

In Sections \ref{subsec:universal_gate_set}, \ref{subsec:distillation} and \ref{subsec:drawbacks}, we examined the conventional Clifford+T distillation-based approach for achieving universal fault-tolerant quantum computation (FTQC) and identified its significant drawbacks, particularly in the EFT era. We refer to this approach as \texttt{qec-conventional}. To address these limitations, we explore two key methods: 

\begin{enumerate}
    \item \textit{Avoid decomposing non-Clifford gates into Clifford+T:}  
   In variational quantum algorithms (VQAs), the primary non-Clifford operations encountered are rotations about the $z$-axis, i.e., \( R_{z}(\theta) \). By retaining these rotations in their native form, we eliminate the additional gate overhead associated with decomposing them into Clifford+T gates.
   \item \textit{Leverage magic state injection for \( R_{z}(\theta) \) rotations:} As discussed in Section \ref{subsec:injection}, we employ the technique introduced by Lao et al. \cite{lao2022magic} to generate \( R_{z}(\theta) \) magic states. Once the \( R_{z}(\theta) \) magic state is prepared, the circuit shown in Figure~\ref{fig:rotation}(C) is used to consume the magic state and perform the desired rotation. This process is probabilistic: If the measurement outcome is \( 0 \), the \( R_{z}(\theta) \) rotation is successfully applied. However, if the outcome is \( 1 \), the result is an unintended rotation of \( R_{z}(-\theta) \). To correct this, a compensatory \( R_{z}(2\theta) \) rotation is applied. This process is repeated in a repeat-until-success fashion until the output bit is \( 0 \). Consequently, a circuit resembling Figure \ref{fig:rotation}(A) before execution may dynamically resemble Figure \ref{fig:rotation}(B) at runtime.
\end{enumerate}

While this approach may not achieve logical error rates as low as \texttt{qec-conventional} in regimes with millions of physical qubits and very low physical error rates, it is far more effective in the resource-constrained EFT era, as we demonstrate in subsequent sections. This approach of error correction only for Clifford gates has been explored in previous work \cite{akahoshi2024partially}, and we adapt it to VQAs. In the context of this paper, we term this approach \textit{partial Quantum Error Correction} (pQEC).



\subsection{pQEC vs \texttt{qec-conventional} for EFT}\label{subsec:pqec_vs_qec_conv}

Figure \ref{fig:pQEC_vs_qec_conventional} illustrates the relative fidelity improvement of variational quantum algorithms (VQAs) when executed using pQEC compared to \texttt{qec-conventional} in the EFT regime. The VQA circuit sizes range from 12 to 24 qubits. \revision{The VQA structure is a fully-connected hardware efficient ansatz with depth$(p)=1$ \cite{kandala2017hardware}}. For \texttt{qec-conventional}, we use 15-to-1 magic state factories with four configurations of \( d_{x}, d_{z}, d_{m} \), chosen due to their compatibility with the 10000 physical qubit constraint \cite{litinski2019magic} (for details on distillation and factories see Section \ref{subsec:distillation}). Note that the 24-qubit VQA circuit paired with the \((15\text{-to-1})_{17, 7, 7}\) factory exceeds this limit by 400 qubits, but the data point is included for completeness. Across all configurations, pQEC consistently outperforms \texttt{qec-conventional}, for several reasons:

\begin{enumerate}
    \item \textit{Small factories lead to inadequate error correction:}   
   For smaller configurations like \((15\text{-to-1})_{7, 3, 3}\), the logical error rate of the distilled T gate is significantly higher than tolerable. Specifically, this factory produces T gates with an error rate of \( 5.4 \times 10^{-4} \) for a physical error rate of \( 10^{-3} \) --- a negligible improvement. This marginal gain is overwhelmed by the substantial increase in gates introduced through Gridsynth decomposition into the Clifford+T gate set, along with the associated errors. Consequently, the overall fidelity is markedly lower compared to pQEC.

   \item \textit{Large factories incur excessive memory errors:}  
   Larger configurations like \((15\text{-to-1})_{17, 7, 7}\) achieve a T gate logical error rate of approximately \( 10^{-8} \), which is extremely low. However, these factories take significant time to distill one T state --- 42 clock cycles in this case. This extended duration forces the program to stall, leading to a considerable accumulation of memory errors. These errors can degrade the program fidelity to levels below that of pQEC, even when the T gates themselves are error-free.

   \item \textit{Optimal factory configurations are still outperformed by pQEC:} There appears to be a balance point for magic state factories, exemplified here by \((15\text{-to-1})_{11, 5, 5}\), which produces T states with sufficiently low error rates at a reasonably high frequency. This configuration achieves lower gate and memory errors than others. However, even in this ``sweet spot'', pQEC outperforms \texttt{qec-conventional} by a factor of 1-2.5x (see inset figure). Furthermore, as the number of qubits in the circuit increases, the relative advantage of pQEC over \texttt{qec-conventional} grows monotonically. This trend strongly suggests that even the most effective configurations for \texttt{qec-conventional} degrade significantly for larger, realistic quantum applications that require \( >100 \) logical qubits.
\end{enumerate}

\begin{figure}[h!]
    \centering
    \includegraphics[width=0.95\linewidth]{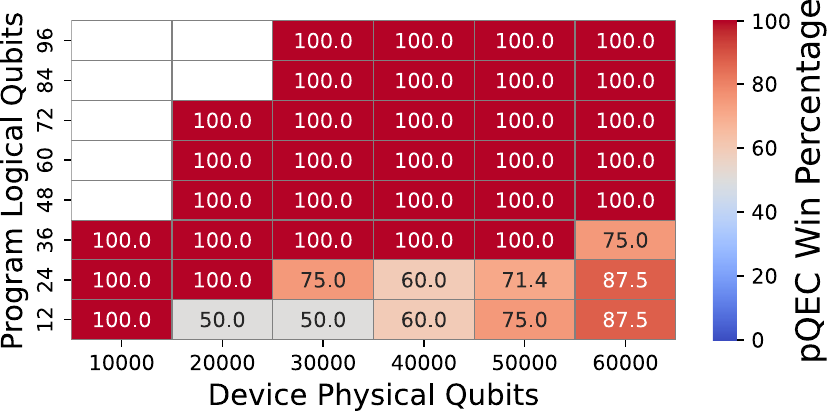}
    \caption{\revision{Win percentage of pQEC over \texttt{qec-conventional} for programs and devices of various sizes. A white square at location $(x, y)$ means that a program with $y$ logical qubits does not fit in a device with $x$ physical qubits for code distance $d=11$.}}
    \label{fig:scaling_study}
\end{figure}



\subsection{\revision{Scaling pQEC beyond 10k qubits}}\label{subsec:scalability}
\revision{In Section \ref{subsec:pqec_vs_qec_conv}, we characterize the EFT era with 10000 physical qubits. However, pQEC remains valid for larger systems. Figure \ref{fig:scaling_study} shows the win percentage of pQEC over \texttt{qec-conventional} for VQAs on different system sizes of 10000 \dots 60000 physical qubits. As the system size increases, \texttt{qec-conventional} outperforms pQEC for VQAs with relatively smaller number of logical qubits. This is because there is increased space available to fit even bulky distillation factories which have a large space-time overhead and produce high-fidelity T states. Since we can fit many such factories, the effective rate of production of T states increases and memory errors do not accumulate as much. A greater number of physical qubits allows using codes with a larger code distance, which also reduces memory errors.} 

\revision{However, as the program size increases and physical qubit constraints are re-established, pQEC 
outperforms
\texttt{qec-conventional}. Thus, in the long run, if the system size is much greater than the resources required to run a bare-bones version of a quantum circuit, \texttt{qec-conventional} is the better strategy. However, if we are operating at the frontier of device capabilities, pQEC wins, demonstrating it usability for large system sizes.}

\subsection{\revision{pQEC versus magic state cultivation}}\label{subsec:cultivation}

\begin{figure}[h!]
    \centering
    \includegraphics[width=\linewidth]{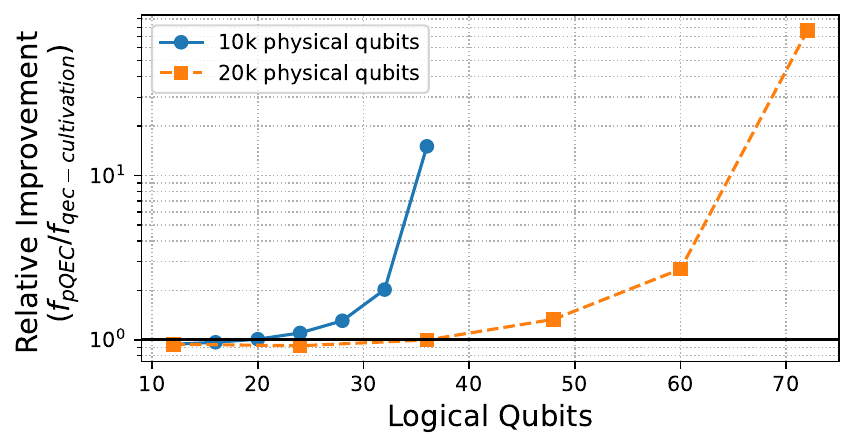}
    \caption{\revision{Relative fidelity improvement for VQAs when executed using pQEC compared to \texttt{qec-cultivation}. For all cases, pQEC does as good as or outperforms \texttt{qec-cultivation}.}}
    \label{fig:cultivation_comparison}
\end{figure}

\revision{Concurrent to the submission of this work, a technique called \textit{magic state cultivation} (MSC) was proposed~\cite{gidney2409magic}, which creates high-fidelity T states with a substantially lower overhead than any T state factory. It has a space overhead comparable to that of a single surface code patch, but can have a high discard rate resulting in a large temporal overhead. Although not commonly used in the literature yet, MSC is seen as a paradigm-shifting way to produce high-fidelity T states. We refer to this approach as \texttt{qec-cultivation}. It involves decomposing non-Clifford gates into Clifford+T operations and then generating T states using MSC (instead of distillation). }

\revision{Figure \ref{fig:cultivation_comparison} shows the relative fidelity improvement of pQEC over \texttt{qec-cultivation} for devices with a 10000 or 20000 physical qubit constraint. For a small number of logical qubits, \texttt{qec-cultivation} beats pQEC. However, as the number of logical qubits in the program increases, and consequently the number of MSC units reduce, the time to generate one T state increases. The remaining qubits stall and accumulate memory errors in that time which reduces the overall fidelity. Considering that the eventual goal is to execute practical applications, it is intuitive that more logical qubits are always better and the benefits achieved by our proposed method in this regime are clearly evident.}
\section{Circuit execution and architecture optimization}\label{sec:circ_optimizations}

When executing a quantum circuit, especially in environments with constrained resources, minimizing resource overheads becomes essential. This section outlines several strategies to address this challenge. To effectively evaluate resource efficiency, the requirements of a quantum circuit are measured using the following metrics:

\begin{enumerate}
    \item For one operation, \textit{space} is equal to the number of physical qubits engaging in the operation: $N_{op} = \left|\{q_{op}\}\right|$. For a circuit, it is the number of distinct physical qubits used across operations in the circuit: $N_{circ}=\left|\bigcup\limits_{op \in circ}\{q_{op}\}\right|$.
    \item For an operation, the \textit{time} $t_{op}$ is equal to the number of clock cycles it takes to complete the operation. For the entire circuit, it is equal to the total time taken by operations along the critical path, denoted $t_{circ}=\sum\limits_{op \in critical\_path}t_{op}$.
    \item \textit{Spacetime volume} quantifies the resource overhead required for executing the circuit or an operation. For a particular operation $V_{op}$, it is equal to $t_{op} \times N_{op}$. A circuit's spacetime volume is the sum of the spacetime volume of each operation: $V_{circ} = \sum\limits_{i} V_{op_{i}}$.
\end{enumerate}

In the error-corrected regime, it is often feasible to balance space and time costs, enabling tailored optimization based on specific constraints. Spacetime volume serves as a comprehensive metric, capturing both spatial and temporal overheads within a unified framework for resource evaluation \cite{litinski2018lattice, gidney2021factor, litinski2019magic}. Within this context, we propose the following optimizations.

\subsection{Efficient layout}\label{subsec:good_layout}

Since the EFT era is constrained by a limited number of physical qubits (assumed to be 10000 in this work), it is essential to design a logical qubit layout that minimizes the number of ancillae to maximize the size of executable quantum programs. We define the \textit{packing efficiency} (PE) as the ratio of data qubit patches to total patches. A layout with a high PE is desirable. However, simply packing the grid predominantly with data qubits increases space and time overhead caused by rotation and alignment. \cite{litinski2019game, geher2024error, tan2024sat}. Additionally, such a layout could restrict interactions between many pairs of data qubits, as the lack of continuous ancilla space between them may hinder lattice surgery operations \cite{litinski2018lattice}. \revision{Thus, we need a layout that gives us a high packing efficiency while also ensuring fast execution of the program.}

The proposed layout, illustrated in Figure \ref{fig:tile_layout}, addresses these challenges. This layout is parameterized by \( k \), which determines the number of logical qubits available for executing a variational quantum algorithm (VQA) circuit. \revision{The yellow patches represent data qubits and the blue patches denote ancilla qubits. The ancilla qubits may either be used as routing space or for injecting \( R_z(\theta) \) magic states (the shaded blue patches).}  It is worth noting that a single magic state (one shaded patch) can theoretically be used to apply rotations on multiple data qubits concurrently \cite{litinski2019magic}, provided the rotation angle \( \theta \) is identical for all those data qubits. However, in typical VQA circuits, the rotational gates on different data qubits often require different rotation angles. Consequently, multiple magic state patches are usually necessary. \revision{The proposed layout enables the consumption of upto $2\lfloor \frac{k}{3} \rfloor$ magic states in parallel. }

For our layout, the packing efficiency is given by  
\[
\text{PE} = \frac{4 \cdot (k+1)}{6 \cdot (k+2)}.
\]  
For large \( k \), this results in a packing efficiency of approximately \( 67\% \). 

\begin{table}[]
    \centering
    \begin{tabular}{|c|c|c|c|}
    \hline
         Layout & linear & fully\_connected & blocked\_all\_to\_all \\
         \hline
         Compact & 1.04 & 1.02 & 1.81 \\
         \hline
         Intermediate & 1.19 & 1.15 & 1.93 \\
         \hline
         Fast & 2.7 & 2.6 & 4.06 \\
         \hline
         Grid & 5.3 & 5.08 & 7.92 \\
         \hline
    \end{tabular}
    \caption{\revision{Spacetime volume of VQAs on different standard layouts relative to our proposed layout.}}
    \label{tab:layout_stv_comparison}
\end{table}

\revision{We evaluate the usefulness of our layout compared to other standard layouts by comparing the space-time volumes ($V_{circ}$) of the executed VQA. We use $V_{circ}$ since it accounts for both space and time effectiveness. By over-provisioning for space, we can reduce execution time and vice versa. We compare our layout against the three layouts Compact, Intermediate and Fast given in Ref.~\cite{litinski2019game} and a Grid layout \cite{Javadi_Abhari_2017}. We consider three different types of VQE ansatze --- Linear Hardware-Efficient Ansatz, Fully-Connected Hardware-Efficient Ansatz \cite{kandala2017hardware} and the \texttt{blocked\_all\_to\_all} ansatz that we describe in detail in Section \ref{subsec:layout_aware_ansatz}.} 

\revision{We take multiple instances of each ansatz type ranging from 8 to 164 qubits (at intervals of 4 qubits) and compute the ratio $V_{circ}(x)/V_{circ}(our\_layout)$ for the ansatz on a layout $x$. Table \ref{tab:layout_stv_comparison} shows the resulting space-time volume ratios. We see that all ratios are greater than or equal to one, suggesting that our layout minimizes space-time volume for all ansatze.}

\revision{Note that these numbers are indicative of the performance of these layouts for VQAs only. All popular VQA ansatze have a serial CNOT ladder structure which introduces data dependencies and minimizes opportunity for parallelism \cite{sim2019expressibility}. The only potential for parallelism exists in the execution of consecutive CNOT gates all of which have the same control qubit. However, we already exploit this in our layout as shown in Figure \ref{fig:rotation_cost}. This algorithmic serialization means that layouts like Fast or Grid are not able to utilize the additional space available for parallelization and result in high space-time volumes as in Table \ref{tab:layout_stv_comparison}. Our layout makes use of this algorithmic serialization and provisions for just enough ancilla space. This keeps the packing efficiency high while being able to exploit maximum opportunity for parallelism. However, for a general program this might not be the optimal layout as there may be more opportunity for parallelization there.}


\revision{}

\begin{figure}[ht!]
    \centering
    \includegraphics[width=0.5\linewidth]{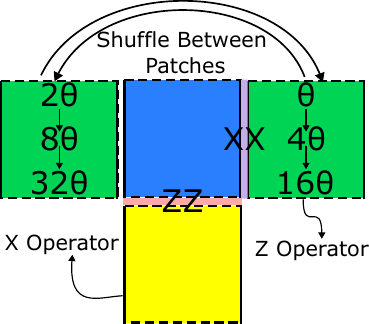}
    \caption{In the time it takes for the ZZ and XX measurement operations to complete their operations with angle $\theta$ in the right green patch, we can inject angle $2\theta$ in the left green patch. If the measurement output post consumption (refer to Figure \ref{fig:rotation}(C)) is 1, we redo the consumption with the injected $2\theta$ patch on the left. In the meantime, we inject an angle of $4\theta$ on the left green patch. This is repeated until the consumption is successful (measurement output 0).}
    \label{fig:patch_shuffling}
\end{figure}


\begin{figure}[ht!]
    \centering
    \includegraphics[width=\linewidth]{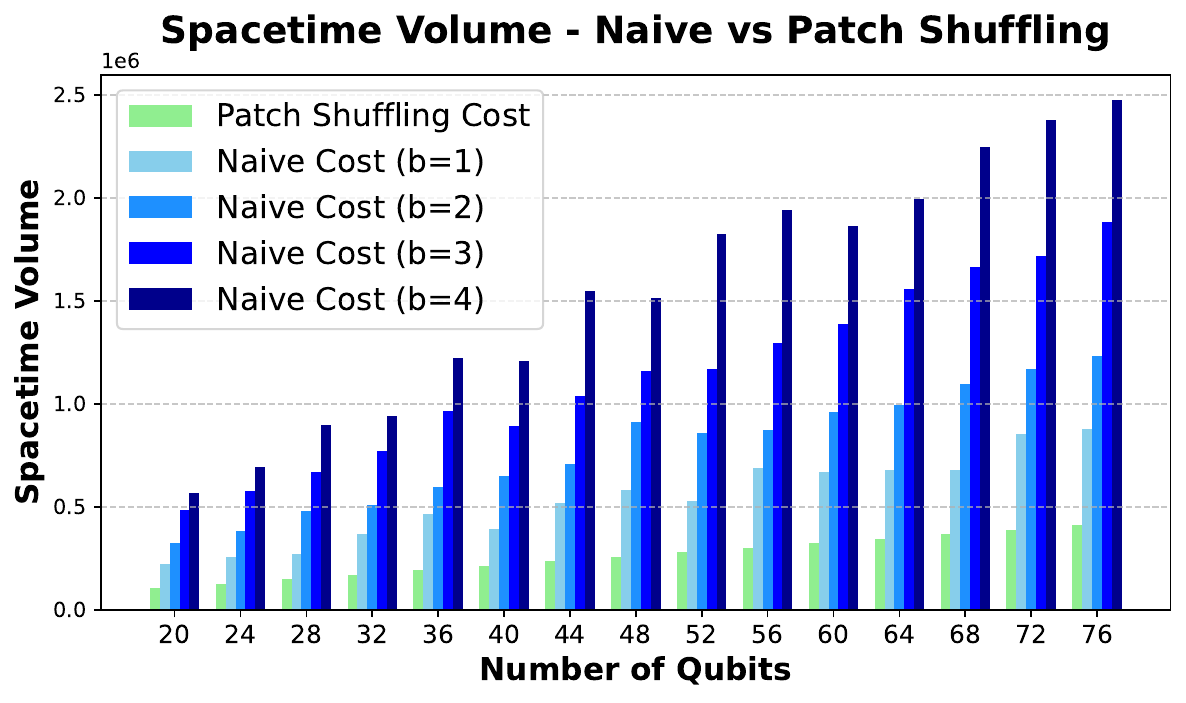}
    \caption{The space-time volume for patch shuffling is significantly lower than that of the naive approach for different number of backup states (``b''). We achieve the lowest space-time volume while keeping memory errors the lowest.}
    \label{fig:patch_shuffling_costs}
\end{figure}

\subsection{Patch shuffling for state injection}\label{subsec:patch_shuffling}

\begin{figure*}[h!]
    \centering
    \includegraphics[width=\linewidth]{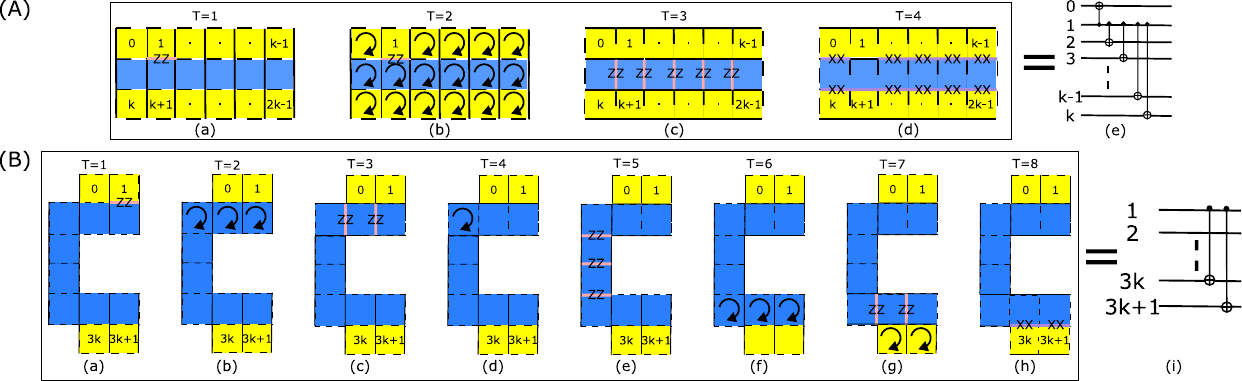}
    \caption{ This Figure uses qubit numbering from Figure \ref{fig:tile_layout}. (A) Sequence of lattice surgery and rotation operations needed to perform a single-control multi-target CNOT between qubit $1$ (control) and qubits $0, 2, 3...2k-1$ (targets). The time taken is 4 code cycles according to the operations specified in \cite{litinski2018lattice, tan2024sat}. (B) Sequence of operations needed to perform CNOTs between qubit $1$ (control) and qubits $3k, 3k+1$ which takes 8 code cycles. Thus, all CNOTs are not of equal cost. One could potentially discard CNOTs that increase circuit duration without substantial contribution to ansatz expressivity.}
    \label{fig:rotation_cost}
\end{figure*}

We previously discussed how the process of magic state injection, followed by consumption (example execution in Figure \ref{fig:rotation}(C)), can fail with a \(50\%\) probability. Consequently, to successfully implement a rotation of \(\theta\), compensatory magic states may need to be injected. For instance, if the injection of a \(\theta\) magic state fails, a compensatory magic state with an angle of \(2 \theta\) must be injected --- this occurs with \(50\%\) probability. If both \(\theta\) and \(2 \theta\) injections fail, a compensatory state of \(4 \theta\) must be injected (\(25\%\) probability), and so on (see Figure \ref{fig:rotation}(B)).

While one approach to implement this is to wait for a consumption step to fail before injecting a new compensatory state, this can lead to additional memory errors due to delays. To mitigate this, a naive strategy involves preparing all compensatory magic states at the start of computation. Based on our analysis, ensuring the availability of magic states up to \(8 \theta\) (4 magic states in total) eliminates stalls with a \(93.75\%\) probability. However, this approach significantly reduces layout utilization in the common case, introduces crowding, and complicates qubit placement. Moreover, the additional magic state patches and corresponding ancilla routes connecting them to data qubits increase both space overhead and the spacetime volume of the circuit.

To address these challenges, we propose a method called \textit{patch shuffling}, which uses only two magic state patches. The process begins with injecting the first patch with an angle of \(\theta\) and the second with \(2 \theta\). The first magic state is consumed by the data qubit using the consumption circuit. If this step fails (measurement output 1), the \(2 \theta\) magic state patch is consumed. Simultaneously, while the \(2 \theta\) patch is being consumed, the first patch is re-injected with \(4 \theta\) as a backup. This ensures that if the second consumption step fails, the \(4 \theta\) patch is ready for immediate use. 

This approach works because, for a surface code patch of distance \(d\), the consumption time equals \(2 d\) clock cycles. Our analysis shows that for a physical error rate of \(10^{-3}\) (typical for EFT), the injection process can complete in less than \(2 d\) cycles with very high probability. Thus, by the time the \(2 \theta\) patch consumption step concludes, the \(4 \theta\) patch is ready for use. This process continues iteratively (with \(8 \theta\) and so on)  until the consumption step succeeds.

The term ``with very high probability'' reflects the inherent characteristics of the magic state injection procedure defined in Ref. \cite{lao2022magic}, which requires two rounds of post-selection. This repeat-until-success procedure can theoretically run for an infinite number of cycles with exponentially small probability, though practical implementations require far fewer cycles.

Figure \ref{fig:patch_shuffling} illustrates how patch shuffling alternates between two patches to implement the desired rotation. Figure \ref{fig:patch_shuffling_costs} compares the spacetime volume of patch shuffling with the naive strategy for varying numbers of backup states (denoted by ``b''). Patch shuffling achieves the optimal balance of low spacetime volume and zero stalls, whereas the naive strategy's spacetime volume increases with the number of backup states, despite reducing stalls.


\subsection{Layout-aware ansatz design}\label{subsec:layout_aware_ansatz}

In Section \ref{subsec:nisq_comparison}, we demonstrate that pQEC achieves better fidelity than NISQ for the fully connected hardware-efficient ansatz. While the layout shown in Figure \ref{fig:tile_layout} can implement the fully-connected ansatz, it is not ideally suited for it. This is because hardware-efficient ansatze commonly used in the NISQ era are primarily motivated by NISQ hardware constraints and the need to minimize circuit depth and gate count to fit within limited NISQ resources. 

In the EFT era, qubit layouts are constrained by small QEC code distances, providing only a limited error correction budget. Memory errors, which arise from delays on the critical path, can quickly deplete this budget and must be minimized. Furthermore, slow operations increase the circuit's execution time (and space-time volume), necessitating careful optimization to execute as many operations as possible while keeping latency low. Thus, these constraints must be accounted for in designing an EFT ansatz.

To design an efficient ansatz, we first examine the cost of implementing entangling gates. Figure \ref{fig:rotation_cost}(A) illustrates CNOT operations between qubit \(1\) (control) and qubits \(0, 2, \ldots, 2k-1\) (targets), which take a total of 4 clock cycles. In each clock cycle, we perform one of three steps: (a) XX measurement, (b) ZZ measurement (both of which are needed to perform CNOTs via lattice surgery \cite{fowler2018low}), or (c) patch rotation such that the correct edges are aligned to perform the lattice surgery operations \cite{tan2024sat}. Note that multiple CNOTs controlled by the same qubit incur the same cost as a single CNOT \cite{litinski2019magic}, allowing \(2k-1\) CNOT gates to be implemented within just 4 cycles. However, Figure \ref{fig:rotation_cost}(B) shows that performing CNOTs between qubit \(1\) (control) and qubits \(3k\) and \(3k+1\) requires 8 clock cycles --- double the latency. This additional overhead arises from extra patch rotation operations necessary to align the correct operator edges during lattice surgery. As a result, program dependencies on these slower CNOTs force subsequent operations to stall longer, and ancilla space remains occupied for an extended period, further delaying other operations.

\begin{figure}[h!]
    \centering
    \includegraphics[width=0.8\linewidth]{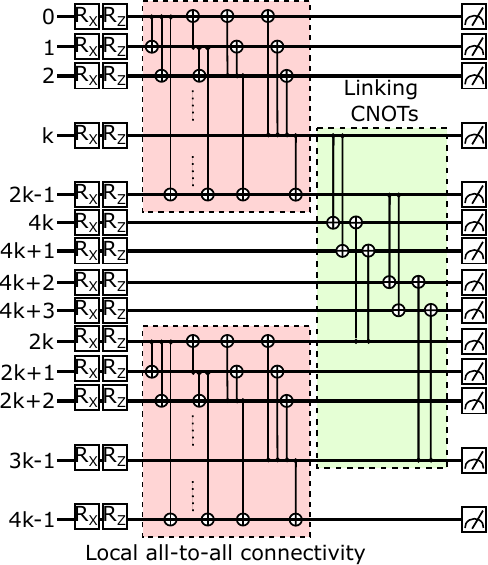}
    \caption{Our proposed ansatz \texttt{blocked\_all\_to\_all}. As the name suggests, there is all-to-all connectivity inside each sub-circuit or ``block'' (pink). Each block is comprised of $2k$ clusters of CNOTs, each of which takes 4 cycles as shown in Figure  \ref{fig:rotation_cost}. The blocks are connected using 8 (fixed number) ``linking CNOTs'' (green) which take a total of $4\times8=32$ cycles to execute. The qubit numbering is as in Figure \ref{fig:tile_layout}.}
    \label{fig:improved_ansatz}
\end{figure}

Building on these insights, we propose a new ansatz called \texttt{blocked\_all\_to\_all}, which is significantly faster compared to the fully-connected hardware-efficient (FCHE) ansatz. The advantage comes from emphasizing logical CNOTs that are fast to implement while minimizing the use of slower CNOTs, reducing both stalls and memory errors. Figure \ref{fig:improved_ansatz} illustrates this ansatz, featuring local blocks of fast, all-to-all CNOTs connected by a sparse number of slow ``linking'' CNOTs between blocks. This structure maintains high connectivity within blocks while reducing the overall latency and overhead. Table \ref{tab:time} shows a comparison of clock cycles taken by each ansatz for different number of qubits. Notably, if the \texttt{blocked\_all\_to\_all} ansatz achieves similar expressivity to the FCHE ansatz for a given problem, it will deliver much higher execution fidelity due to its lower space-time volume and reduced error rates.

\begin{table}[h!]
\centering
\caption{Cycles taken by \texttt{blocked\_all\_to\_all} vs FCHE ansatz.}
\begin{tabular}{|c|c|c|c|}
\hline
\textbf{Qubits} & \textbf{20} & \textbf{40} & \textbf{60} \\ \hline
\texttt{blocked\_all\_to\_all} & 71 & 121 & 171 \\ \hline
 FCHE & 131 & 271 & 411 \\ \hline
\end{tabular}
\label{tab:time}
\end{table}

Finally, we emphasize that this proposed ansatz is primarily intended to demonstrate the importance of designing EFT-tailored ansatze with high expressivity and low latency, akin to the NISQ era. However, creating an optimal ansatz requires extensive analysis of application-hardware trade-offs, varying across different use cases and architectures. This remains a complex and expansive research area beyond the scope of this work.


\subsection{EFT ansatz design: $R_z(\theta)$ to CNOT ratio}\label{subsec:rz_vs_cnot_ratio}

When designing a general ansatz for the EFT regime, it is crucial to ensure that the resulting circuit consistently performs better with pQEC compared to NISQ across reasonable values of \( N \) (number of qubits) and \( p \) (circuit depth). In our EFT framework, for large \( p \), the ratio of \( R_z(\theta) \) to CNOT gates in the ansatz becomes a critical metric for predicting this success.

For NISQ execution, we consider the following error rates: CNOT error \( p_{phys} \), non-\( R_z \) single-qubit gates error \( p_{phys}/10 \), \( R_z(\theta) \) error 0, and measurement error \( 10 p_{phys} \), with $p_{phys}=10^{-3}$ \cite{mckay2017efficient}. These values are standard in prior works \cite{lao2022magic} and align with experimental data \cite{ibm_machines}. In VQAs, non-\( R_z \) single-qubit gates are rare, and the number of measurement operations does not scale with circuit depth. Therefore, CNOT gates dominate NISQ errors, with error contributions scaling with both circuit width and depth.

With pQEC, all single-qubit Clifford gates, measurements, and CNOT gates are error-corrected. For \( d=11 \) surface codes and \( p_{phys}=10^{-3} \), the error rates for memory, measurement, CNOT, and single-qubit Clifford gates are all approximately \( 10^{-7} \). However, \( R_z(\theta) \) gates are injected and not error-corrected, maintaining an error rate of \( \sim 10^{-3} \), making \( R_z(\theta) \) gates the dominant error source.

Note that for smaller values of \( p \) or \( d \), additional error sources may need consideration. For example, measurement noise in the NISQ regime or Clifford gate errors in the pQEC regime could contribute significantly.
Overall, the ratio of dominant error sources in each regime primarily determines whether pQEC outperforms NISQ. For practical application sizes, this determination hinges on the relative contributions of CNOT and \( R_z(\theta) \) errors. For a CNOT error rate of \( 10^{-3} \) in the NISQ regime, and an \( R_z(\theta) \) error rate of \( 0.76 \times 10^{-3} \) in the EFT regime, theoretically, an ansatz will perform better using pQEC if the number of CNOTs grows faster than $0.76$ times the number of \( R_z(\theta) \) gates. 



\begin{figure}[h!]
    \centering
    \includegraphics[width=0.95\linewidth]{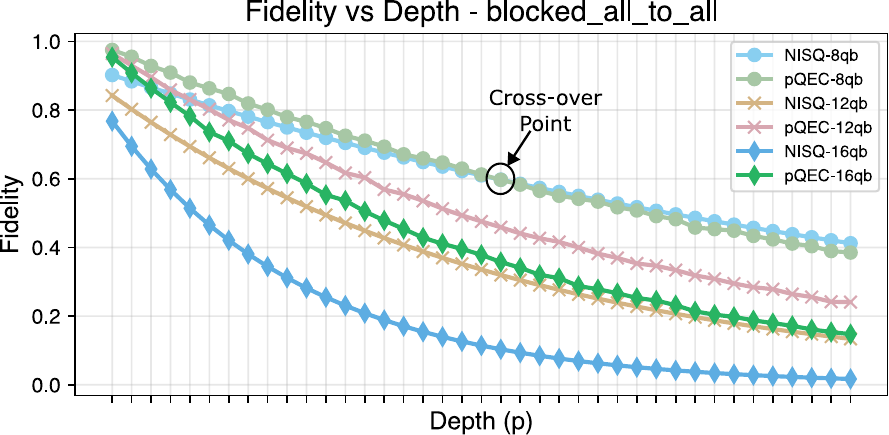}
    \caption{Fideilty of the different sized \texttt{blocked\_all\_to\_all} ansatze evaluated in the NISQ and EFT regimes. We see that in case of \texttt{blocked\_all\_to\_all}, for 8 qubits, the NISQ fidelity exceeds the EFT fidelity at higher depth. However,  for 12 and 16 qubits there is a consistent benefit of pQEC over NISQ.}
    \label{fig:esp_vs_depth}
\end{figure}

For the \texttt{blocked\_all\_to\_all} ansatz, the number of $R_{z}(\theta)$ gates is $2 N p \cdot \mathbb{E}[g]$, where $\mathbb{E}[g]$ is the expected number of $R_{z}(\theta)$ gates required at runtime for each logical $R_{z}(\theta)$ gate in the circuit. For our proposed scheme, $\mathbb{E}[g]=2$ (expectation value of geometric series with $p_{succ}=p_{fail}=0.5$). The number of CNOTs is $(\frac{N^{2}}{2}-5N+20) p$. Consequently, the ratio of CNOTs to $R_{z}(\theta)$ is $\frac{N}{8} - \frac{5}{4} + \frac{5}{N}$, which exceeds $0.76$ for all $N \geq 13$. This means that for large depths, pQEC should provide benefits for problem sizes exceeding 13 qubits, covering essentially all useful problem instances. Experimental results in Figure \ref{fig:esp_vs_depth} validate this theory with slight deviations.

Figure \ref{fig:esp_vs_depth} illustrates the fidelity of \texttt{blocked\_all\_to\_all} ansatz in both NISQ and pQEC regimes for varying qubit sizes. For 8 qubits, the rate of decay in NISQ fidelity is slower than in the EFT fidelity which means that for large depths, NISQ is a better regime for operation. However, for 12 and 16 qubits, the inverse trend emerges, clearly demonstrating the growing advantage of EFT over NISQ as problem sizes increase. Note that practically, the crossover point is observed around 12 qubits, a little before the theoretically estimated value of 13. However, the trends are same.

For the UCCSD and FCHE ansatze, the CNOT-to-\( R_{z} \) ratio scales as \( \mathcal{O}(N) \). This indicates that at larger depths, these ansatze will also perform better in the EFT regime compared to NISQ. Note that the result for the UCCSD ansatz would look similar because of the same CNOT-to-\( R_{z} \) ratio. Even without accounting for their prohibitively large depths, the UCCSD ansatze are inherently better suited for execution in the EFT regime compared to NISQ simply based on the relative number of CNOT and \( R_{z}(\theta) \) gates.



A commonly used ansatz in the NISQ era is the linear ansatz. For a linear ansatz of $N$ qubits and depth $p$, the number of CNOTs is $N p$ while the number of $R_{z}(\theta)$ gates is $2 N p \cdot \mathbb{E}[g]$. For this configuration, the growth rate of CNOT count is only $0.25$x the rate of growth of $R_{z}(\theta)$ count, which is much lower than $0.76$x. Consequently, a linear ansatz is not a good choice for the pQEC regime. However, this limitation is not particularly consequential, as the hardware-efficient ansatz, including the linear ansatz, is generally considered a suboptimal choice for experiments with meaningful practical applications.

\section{Methodology}\label{sec:methodology}

\begin{figure*}[h!]
    \centering
    \includegraphics[width=0.8\linewidth]{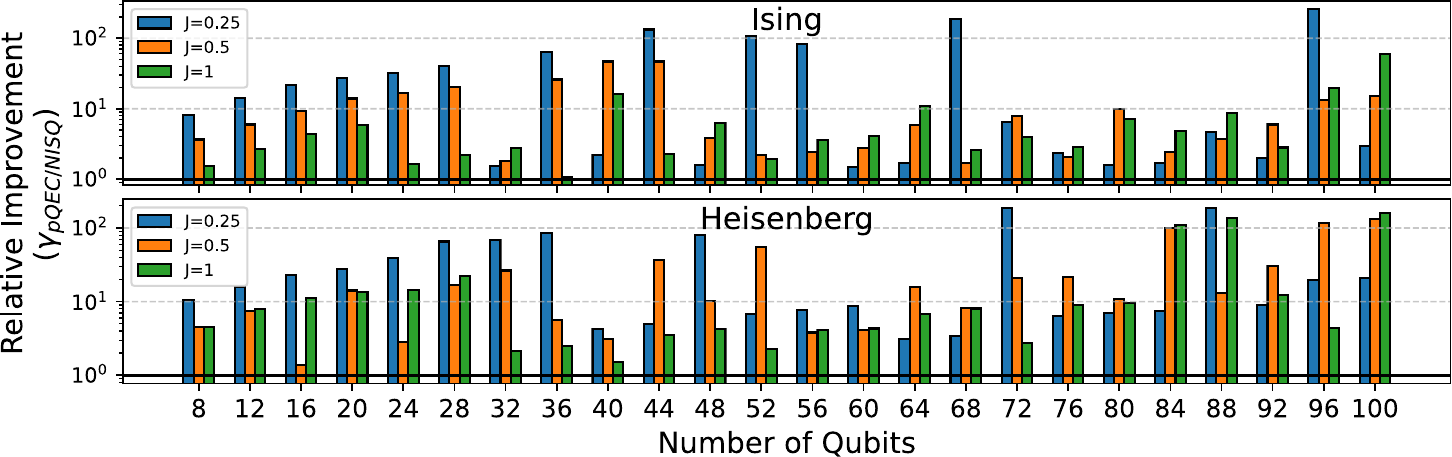}
    \caption{Relative improvement $\gamma$ of pQEC compared to NISQ for the Ising and Heisenberg models. \revision{$\gamma_{average}$(Ising) = 6.833x}, $\gamma_{max}$(Ising) = 257.54x. \revision{$\gamma_{average}$(Heisenberg) = 12.59x}, $\gamma_{max}$(Heisenberg) = 189.54x.}
    \label{fig:gamma_pqec_vs_nisq_cliff}
\end{figure*}

\begin{figure}[htp!]
    \centering
    \includegraphics[width=0.9\linewidth]{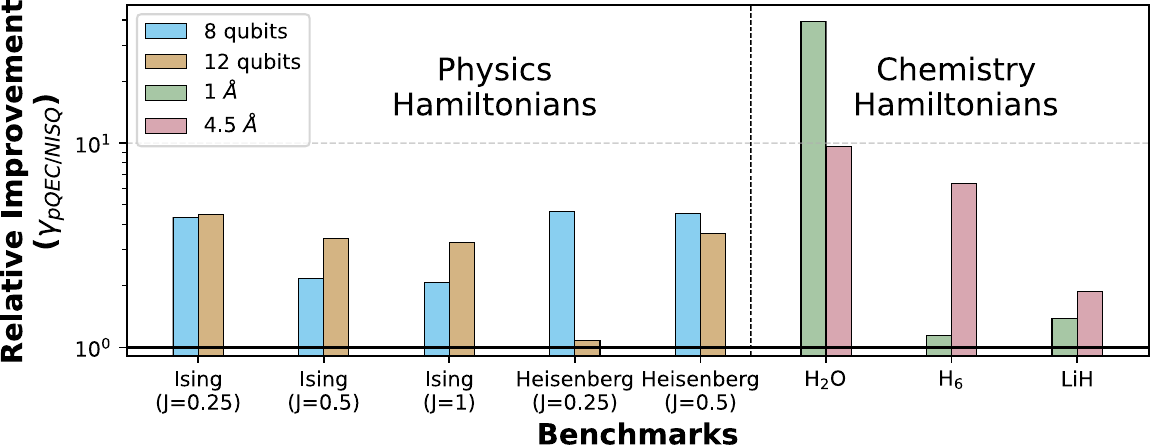}
    \caption{$\gamma_{pQEC/NISQ}$ calculated for a variety of physics and chemistry Hamiltonians for problems of size 8 and 12 qubits via noisy density matrix simulations.}
    \label{fig:full_dm_simulations}
\end{figure}

\subsection{VQA benchmarks}\label{subsec:benchmarks}

To study the effectiveness of EFT-VQA, we study the following physics and chemistry models.

\subsubsection{Physics benchmarks}\label{subsubsec:physics}
 We study two common physics Hamiltonians: the Ising model and the Heisenberg model, that are known to be able to show phase transitions \cite{ising, heisenberg}. In this work, we focus on the case of 1D models with constant couplings. The Ising model is given by
\begin{equation}
    \ham = J \sum_{i=1}^{N-1} X_i X_{i+1} + \sum_{i=1}^N Z_i,
\end{equation}
which describes how $N$ qubits coupled along the $X$ axis with strength $J$ interact with each other when there is a unit strength external magnetic field along the $Z$ axis.

We also look at the field-free Heisenberg model which couples spins in all directions.
\begin{equation}
    \label{eq:xxz}
    \ham = \sum_{i=1}^{N-1} \qty(J X_i X_{i+1} + J Y_i Y_{i+1} + Z_i Z_{i+1}).
\end{equation}
The $ZZ$-interaction occurs with a coupling strength of 1.

In this paper we study the cases $J=0.25, 0.50, 1.00$ for both the Ising model and the Heisenberg model.

\subsubsection{Chemistry benchmarks}\label{subsubsec:chem}
We study the H$_{2}$O, H$_{6}$, and LiH molecules. For each molecule, we limit the active subspace to six orbitals, which leads to 12 qubit Hamiltonians. The molecules are constructed using PySCF \cite{sun2018pyscf} and Qiskit Nature \cite{qiskit2024}. For each molecule, we consider two bond lengths $(l)$. The specific configurations are:

\begin{enumerate}
    \item H$_{2}$O --- $l=$ $1$\AA, $4.5$ \AA \, --- 367 terms
    \item H$_{6}$ --- $l=$ $1$\AA, $4.5$ \AA \, --- 919 terms
    \item LiH --- $l=$ $1$\AA, $4.5$ \AA \, --- 631 terms
\end{enumerate}


\subsection{Evaluation infrastructure}\label{subsec:Method}

\subsubsection{Density matrix simulations}\label{subsubsec:density_matrix_sim}

For 8 and 12 qubits, we conduct density matrix simulations using Qiskit's AerSimulator. In the NISQ regime, gate errors are modeled as a combination of depolarizing and thermal relaxation channels, while measurement errors combine bit-flip and thermal relaxation channels. Idling errors are represented using thermal relaxation channels alone. 
In the pQEC regime, gate and memory errors are modeled as depolarizing channels, and measurement errors are represented as bit-flip errors. The error rates for error-corrected operations in the pQEC regime --- specifically $\{Memory, CX, H, S, Measure\}$ --- are derived by simulating these operations using \textit{Stim}~\cite{gidney2021stim} with a physical error rate of \( p = 10^{-3} \). The error rates for state injection operations \( R_{z} (\theta) \) are calculated using Equation (3) from Ref. \cite{lao2022magic}, applying the biased noise model provided in their Figure 6. We use the Cobyla and ImFil optimizers \cite{powell1994direct, doi:10.1137/1.9781611971903} to minimize the energy value. Each benchmark is executed three to five times for different seeds, and we report the best result for each run.

\subsubsection{Clifford state simulations}\label{subsubec:clifford_state_simulations}
To validate our proposal at scale, we evaluate circuits with sizes up to 100 logical qubits. Conducting full density-matrix noisy simulations on classical hardware is impractical for this regime. Therefore, for circuits ranging from 16 to 100 qubits, we constrain the \( R_{z} (\theta) \) arguments to multiples of \( \frac{\pi}{2} \). This converts the circuit into a Clifford circuit, enabling efficient classical simulation using stabilizer simulators like \textit{Stim}~\cite{gidney2021stim}, as supported by the work of Aaronson and Gottesman~\cite{aaronson2004improved}. 
The effect of noise on Clifford states has been extensively used as a proxy for arbitrary quantum states. For instance, IBM’s quantum utility paper~\cite{kim_evidence_2023a} employs this approach. Similarly, Refs.~\cite{ravi2022cafqa, seifert2024clapton} show that in variational quantum eigensolvers (VQE) using hardware-efficient ansatzes, the energy of the best Clifford state approximates the ground state energy of the Hamiltonian. This further validates the suitability of using Clifford states as proxies for full density-matrix simulations.
Our Clifford state simulations include all noise sources that are classically simulable. Depolarizing and bit-flip errors are handled efficiently via classical methods, while non-Clifford thermal relaxation errors are modeled using Pauli twirling approximations as described in Ref.~\cite{ghosh2012surface}. For optimization over the discrete parameter search space, we employ a genetic algorithm, which allows for efficient parallelization and scalability.


\subsection{Evaluation metrics}\label{subsec:metric}

\begin{figure*}[htp!]
    \centering
    \includegraphics[width=0.9\linewidth]{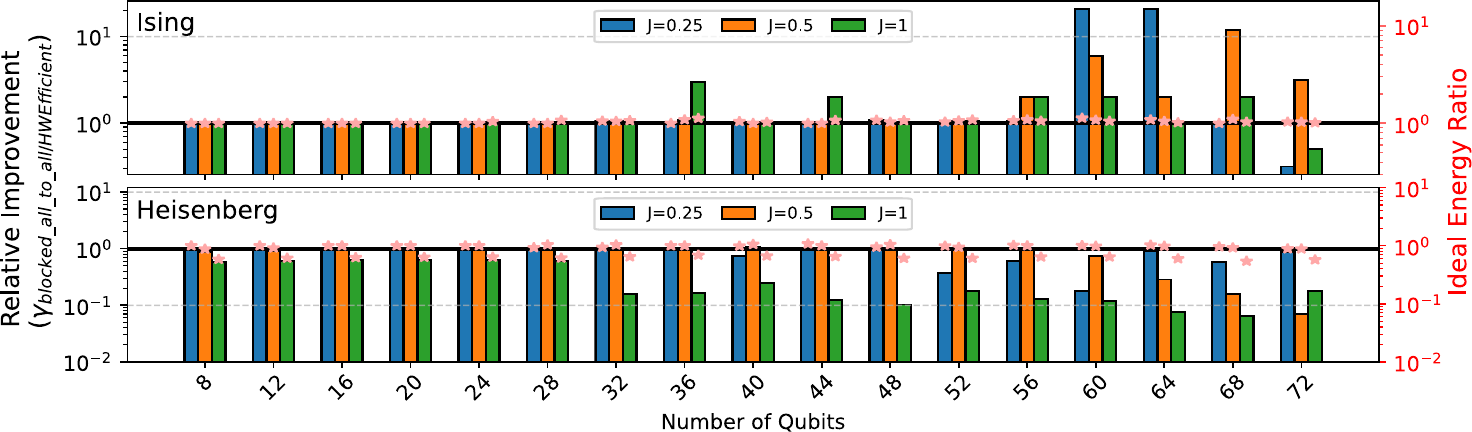}
    \caption{Relative improvement $\gamma$ of \texttt{blocked\_all\_to\_all} compared to Hardware-Efficient Fully-Connected for the Ising and Heisenberg models. \revision{$\gamma_{average}$(Ising) = 1.35x}, \revision{$\gamma_{max}$(Ising) = 21x}. \revision{$\gamma_{average}$(Heisenberg) = 0.49x}, \revision{$\gamma_{max}$(Heisenberg) = 1.05x}}
    \label{fig:blocked_all_to_all_vs_hardware_efficient}
\end{figure*}

\subsubsection{Relative improvement $\gamma_{A/B}$:} \label{subsubsec:relative_improvement}This metric quantifies the extent to which VQE in the regime ``A'' (say pQEC) is able to reduce the gap to some reference energy $(E_{0})$ compared to the VQE in the regime ``B'' (say NISQ). We choose the reference energy $E_{0}$ to be equal to the actual ground state energy for 8 and 12 qubit Hamiltonians. This value is equal to the lowest eigenvalue of the Hamiltonian matrix, which can be obtained by diagonalizing it. For 16+ qubit systems, matrix diagonalization becomes prohibitively expensive on a classical computer. Hence, the value of $E_{0}$ is chosen to be the lowest stabilizer state energy obtained in the absence of noise \cite{seifert2024clapton, ravi2022cafqa}.

\begin{equation}\label{eq:gamm_def}
    \gamma_{A/B} = \frac{E_{0} - E_{B}}{E_{0} - E_{A}}
\end{equation}
\section{Evaluation}\label{sec:evaluation}

\subsection{pQEC versus NISQ}\label{subsec:nisq_comparison}
As discussed in Section \ref{subsec:Method}, we use density matrix simulations for 8, and 12 qubits and Clifford state simulations for larger number of qubits. The VQE ansatz we use is a $depth (p) = 1$ FCHE ansatz. We run our evaluations for all benchmarks described in Section \ref{subsec:benchmarks}.

\subsubsection{Small-scale density matrix simulations} \label{subsubsec:density_matrix_evaluations_nisq_vs_pqec}
Figure \ref{fig:full_dm_simulations} shows the relative improvement $\gamma_{pQEC/NISQ}$ of pQEC over NISQ for Ising, Heisenberg, H$_{2}$O, H$_{6}$, and LiH Hamiltonians. The y-axis is plotted in log scale. We observe the pQEC consistently outperforms NISQ. For the Ising Hamiltonians, the average value of $\gamma$ is \revision{3.45x} and goes up to \revision{4.48x}. For Heisenberg, the average and max $\gamma$ are \revision{3.005x} and 4.64x. The average and max $\gamma$ values for the chemistry Hamiltonians are: H$_{2}$O: average = \revision{19.52x}, max = 39.50x; H$_{6}$: average = \revision{2.69x}, max = 6.34x; LiH: average = \revision{1.61x}, max = 1.88x.

\subsubsection{Large-scale Clifford  simulations}\label{subsubsec:clifford_state_evaluations_nisq_vs_pqec}
Figure \ref{fig:gamma_pqec_vs_nisq_cliff} shows the relative improvement to VQE energies via pQEC, compared to NISQ, for the Ising and Heisenberg Hamiltonians upto 100 logical qubits using Clifford state simulation. We plot the value of $\gamma_{pQEC/NISQ}$ (Equation \ref{eq:gamm_def}) on the $y$-axis, which is in the log scale. We see that for each evaluation, pQEC outperforms NISQ. 
For the Ising model, the average relative improvement of VQE with pQEC compared to NISQ is \revision{$6.83$x} and goes upto $257.54$x. For the Heisenberg model, the average relative improvement of VQE with pQEC compared to NISQ is \revision{$12.59$x} and goes up to $189.54$x. All these results align with our hypothesis that VQE in the presence of pQEC outperforms NISQ substantially at scale.


\subsection{Fully-connected hardware-efficient (FCHE) versus \texttt{blocked\_all\_to\_all} ansatz for pQEC}

Figure \ref{fig:blocked_all_to_all_vs_hardware_efficient} shows the performance of the \texttt{blocked\_all\_to\_all} ansatz relative to the FCHE ansatz in the presence of pQEC. We observe that for the majority of cases, \texttt{blocked\_all\_to\_all} performs comparably to or better than FCHE ansatz. However, there is a non-trivial number of cases where it performs worse. The most noticeable case is Heisenberg ($J=1$), where FCHE consistently performs much better. This is most likely because the ansatz structure does not capture the type of interactions needed to minimize the energy of the Hamiltonian. This shows that the ``optimal'' ansatz varies from one Hamiltonian to the next, as far as minimizing energy is considered. To study the effect of each ansatz's convergence ability independent of noise, we also plot the ratio of ideal energies achieved by each ansatz upon convergence. We see that this ratio hovers around 1 for most benchmarks, indicating similar ``expressibility'' for both ansatz on average. We also see that the cases where \texttt{blocked\_all\_to\_all} performs worse are often the ones for which it has poor expressibility compared to FCHE (ideal energy ratio < 1).

However, we see that \texttt{blocked\_all\_to\_all} on average performs \revision{$1.35x$} better compared to FCHE for Ising models. \revision{For Heisenberg models, it achieves $0.49x$ fidelity of FCHE on average. This average fidelity improvement is reduced primarily because of the $J=1$ case. However, we still universally reduce the time of execution by more than half compared to FCHE (see Section \ref{subsec:layout_aware_ansatz})}.

\section{Discussion and Related Work}\label{sec:related_work}

Recent work on EFT era quantum computing explores better methods for quantum state injection \cite{akahoshi2024partially, toshio2024practical}, proposals to implement partial quantum error correction  \cite{akahoshi2024partially} and application level resource estimation studies \cite{akahoshi2024compilation}. In particular, Akahoshi et. al. \cite{akahoshi2024partially} build upon the work in Ref. \cite{lao2022magic} to propose a better magic state injection strategy and use it for partial error correction. They also propose a general architecture (STAR) and optimizations to leverage their partial quantum error correction strategy. Contrary to the more general approach taken in Ref. \cite{akahoshi2024partially}, our work focuses exclusively on VQAs and proposes application specific spacetime volume optimizations. Our work can be integrated with the optimizations proposed in Ref. \cite{akahoshi2024partially} to further achieve better results.

With EFT systems, limited error correction capabilities makes it beneficial to integrate NISQ-inspired error mitigation methods alongside QEC.
In the NISQ era, various fidelity-boosting techniques have been proposed for VQAs. Although not all of these methods may be suitable for the EFT regime, some may transition effectively. The more obvious examples include pre-processing and post-processing methods like VQA initialization~\cite{lin16,marwaha21,shaydulin21,CAFQA_Ravi2022}, circuit optimizations~\cite{dangwal2023varsaw}, and zero-noise extrapolation (ZNE)~\cite{zne1,zne3,zne4} --- it is intuitive that these will transition because their benefits are mostly independent of how the VQA is executed, although their exact implementation would need to be appropriately modified to be cognizant of QEC and FT computation.
Other techniques such as strategies for managing transient errors~\cite{QISMET_Ravi2022,disq} may be expected to have benefits in the EFT era, however the causes and impact of transient errors can be very different in the EFT space~\cite{acharya2024quantum}.
The EFT utility of some other NISQ-focused VQA techniques are unclear. For example, variational approaches to insert error mitigating gate sequences like dynamical decoupling (DD) into the quantum circuit~\cite{VAQEM_Ravi2022} have a less direct transition to the EFT regime. While it is known that DD sequences improve the fidelity of stabilizer circuits that are integral to QEC~\cite{acharya2024quantum}, optimizing DD sequences for stabilizer circuits is fairly orthogonal to optimizing them for insertion into the VQA circuit. However, it should be noted that the benefit of EM techniques like VAQEM implemented as is cannot directly compete with the benefits of pQEC (this is, of course, not unique to VQAs or to our work). On average, VAQEM provided 2-3x improvements in fidelity on very noisy NISQ machines at $10^{-2}$ two-qubit error rates, whereas pQEC provides 7-13x improvements even over a much improved NISQ baseline with $10^{-3}$ two-qubit error rates.

\begin{figure}[h!]
    \centering
    \includegraphics[width=0.9\linewidth]{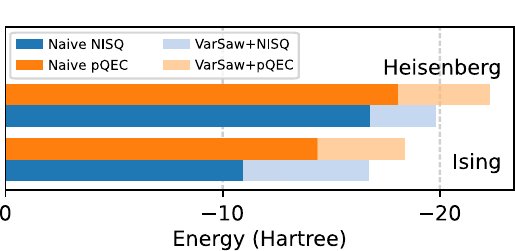}
    \caption{\revision{VarSaw helps VQE converge to a lower energy value for both NISQ and pQEC execution. We demonstrate this for 12 qubit, J=1, Ising and Heisenberg Hamiltonians.}}
    \label{fig:varsaw_effect}
\end{figure}

Implementing and evaluating all of these techniques in the EFT context is beyond the scope of this work. \revision{However, we are able to demonstrate the seamless integration of one of the above techniques, VarSaw with our pQEC framework. Figure \ref{fig:varsaw_effect} shows the improved convergence of VQE to lower energy values in the presence of VarSaw for both NISQ and pQEC execution. We show this for 12 qubit Ising and Heisenberg Hamiltonians. (J=1)}

A critical component of QEC is the decoding process, which identifies and corrects errors. Although the choice of decoder is generally orthogonal to the application, approximate decoders (offering lower classical computational costs) may be particularly attractive in the EFT era due to less stringent error rate requirements. Notable examples include the Union Find decoder \cite{liyanage2023scalable} and pre-decoding methods proposed for long-term FTQC, such as clique-based decoders \cite{Clique_Ravi2022}, Astrea \cite{vittal2023astrea}, and ProMatch \cite{alavisamani2024promatch}.
\section{Conclusion}\label{sec:conclusion}
The Early Fault Tolerance (EFT) era, with 10000-qubit superconducting devices and projected two-qubit error rates of 0.1\%, will enable practical error correction using lightweight codes, significantly improving fidelities over the NISQ era. However, the high qubit overhead of surface codes will limit capabilities, requiring efficient QEC strategies and careful application selection. In this work, we showed that Variational Quantum Algorithms (VQAs) are promising for EFT systems and adapted the ideas of partial quantum error correction (pQEC) to correct Clifford operations while performing \(R_z(\theta\) rotations via magic state injection, avoiding costly T state distillation. Our results highlight that pQEC can significantly boost VQA fidelities, with architectural optimizations further improving fidelity, reducing latency, and enhancing qubit packing efficiency, positioning EFT systems to impact key quantum applications.
\section{\revision{Proof of patch shuffling}}\label{sec:appendix}

\revision{State injection involves the application of a physical gate followed by two rounds of stabilizer measurements. This accomplishes post-selection as well. If the post-selection procedure detects an error, we discard the injected state. The time for application of physical gate is zero cycles. We repeat this process until two subsequent rounds of stabilizer measurements yield all-zero results (successful post-selection). Let $physical\_error\_rate=p, code\_distance=d$. Number of stabilizers $= d^{2}-1$. Denote the probability of event $\mathcal{Z}$ by $\mathcal{P}[\mathcal{Z}]$.}

\revision{$\mathcal{P}$[a single stabilizer detects error]$=2p(1-p)$.\\
$\implies \mathcal{P}$ [at least one stabilizer detects error]$=2p(1-p)(d^{2}-1)$\\
$\implies \mathcal{P}$ [post selection trial fails]$=p_{fail}=2p(1-p)(d^{2}-1)$\\
$\implies \mathcal{P}$ [post selection trial succeeds]=$p_{pass}=1-p_{fail}$\\}
\begin{equation}
    \revision{p_{pass}=1-2p(1-p)(d^{2}-1)}
    \label{eq:p_pass}
\end{equation}

\revision{We keep repeating the trials till we have success. This forms a geometric distribution. Let $\mathcal{X}$ denote the random variable representing the number of trials before success. The Expectation ($\mathcal{E}$) and Standard Deviation ($\sigma$) of this geometric distribution \cite{geometric_rv}, are\\}
\revision{$ \mathcal{E}[\mathcal{X}]=\frac{1}{p_{pass}}$ and $\sigma^{2}[X]=\frac{1-p_{pass}}{p_{pass}^{2}} \implies \sigma[X]=\frac{\sqrt{1-p_{pass}}}{p_{pass}}$.\\}
\revision{Number of trials upto one standard deviation $ = \mathcal{N}_{trials}=\mathcal{E}[\mathcal{X}] + \sigma[X] = \frac{1+\sqrt{1-p_{pass}}}{p_{pass}} = 1.959$ (for $p=10^{-3}$ and $d=11$).\\}
\revision{Probability that we take less than $\mathcal{N}_{trials}$ rounds is $\mathcal{P}[\mathcal{X}<=\mathcal{E}[\mathcal{X}] + \sigma[X]] = 1-(1-p_{pass})^{\frac{1+\sqrt{1-p_{pass}}}{p_{pass}}} = 0.9391.$ This value is called ``high-probability'' in the main text.}

\revision{The time taken to ``consume'' a state according to the circuit in Figure \ref{fig:rotation} (C) is $2d$ cycles (The time to perform a CNOT gate with lattice surgery).\\
We will inject a state with ``high probability" in the same time that another state is being consumed if}
\begin{equation}
    \revision{\mathcal{N}_{trials}=\mathcal{E}[\mathcal{X}] + \sigma[X] \leq 2d}
    \label{eq:high_probability_inequality}
\end{equation}
\revision{From Equation \ref{eq:high_probability_inequality}, we have $\mathcal{E}[\mathcal{X}] + \sigma[X] \leq 2d$\\
$\implies 1 + \sqrt{1-p_{pass}} \leq 2d\cdot p_{pass}$\\
$\implies 4d^{2} \cdot p_{pass} - 4d + 1 \geq 0$\\}
\revision{Using the value of $p_{pass}$ from Equation \ref{eq:p_pass}, we get\\
$4d^{2}(1-2p(1-p)(d^{2}-1))-4d+1 \geq 0$\\
$\implies -8d^{2}(d^{2}-1)p(1-p) \geq -4d^{2}+4d-1$\\
$\implies 8d^{2}(d^{2}-1)p(p-1) \geq -(4d^{2}-4d+1)$\\
$\implies p^{2}-p + \frac{4d^{2}-4d+1}{8d^{2}(d^{2}-1)} \geq 0$}

\revision{Define $c = \frac{4d^{2}-4d+1}{8d^{2}(d^{2}-1)}$, we get the inequality $p^{2}-p+c \geq 0$, which is true for $p \in (-\infty, \alpha] \cup [\beta, \infty)$, where $\alpha=\frac{1-\sqrt{1-4c}}{2}$, and $\beta=\frac{1+\sqrt{1-4c}}{2}$.} 

\revision{For $d=11$, we get $\alpha=0.003811$ and $\beta=0.996189$. Since $p=10^{-3}$, we have $p \in (-\infty, \alpha]$. Hence, the inequality $\mathcal{E}[\mathcal{X}] + \sigma[X] \leq 2d$ is satisfied, and we have $\mathcal{P}[\mathcal{X} \leq 2d] \geq 0.9391$}.
\section{Acknowledgements}\label{sec:acknowledgements}
This work is funded in part by EPiQC, an NSF Expedition
in Computing, under award CCF-1730449; in part by STAQ
under award NSF Phy-1818914/232580; 
in part by the U.S. Department of Energy, Office of Science, Office of Advanced Scientific Computing Research, Accelerated Research in Quantum Computing under Award Number DE-SC0025633;
and in part by the NSF Quantum Leap Challenge Institute for Hybrid Quantum Architectures and Networks (NSF Award 2016136), in part based upon work supported by the U.S. Department of Energy, Office of Science, National Quantum Information Science Research Centers, in part by Wellcome-Leap Q4BIO, and in part by the Army Research Office under Grant Number W911NF-23-1-0077. 
This research used resources of the National Energy Research Scientific Computing Center, a DOE Office of Science User Facility supported by the Office of Science of the U.S. Department of Energy under Contract No. DE-AC02-05CH11231 using NERSC award ASCR-ERCAP0033197. 
The views and conclusions contained in this document are those of the authors and should not be interpreted as representing the official policies, either expressed or implied, of the U.S. Government. The U.S. Government is authorized to reproduce and distribute reprints for Government purposes notwithstanding any copyright notation herein. 
FTC is the Chief Scientist for Quantum Software at Infleqtion and an advisor to Quantum Circuits, Inc. This research used resources of the Oak Ridge Leadership Computing Facility, which is a DOE Office of Science User Facility supported under Contract DE-AC05-00OR22725. 
This work was completed in part with resources provided by the University of Chicago’s Research Computing Center.
We also acknowledge the use of NVIDIA Quantum Cloud resources for preliminary experiments conducted in this research.

\bibliographystyle{ACM-Reference-Format}
\bibliography{sample-base}

\appendix

\end{document}